\documentclass[journal]{IEEEtran}
\usepackage{titletoc}
\usepackage{diagbox}
\usepackage{algorithmic}
\usepackage{makecell}
\usepackage{algorithm}
\usepackage{array}
\usepackage{amsmath,amssymb,amsfonts}
\usepackage[caption=false,font=normalsize,labelfont=sf,textfont=sf]{subfig}
\usepackage{cite}
\usepackage{textcomp}
\usepackage{stfloats}
\usepackage{url}
\usepackage{verbatim}
\usepackage{graphicx}
\usepackage{lipsum}
\usepackage{xcolor}
\usepackage{makecell}
\usepackage{multirow}
\usepackage{amsmath}
\usepackage{color}
\usepackage[colorlinks,
            linkcolor=blue,
            anchorcolor=blue,
            citecolor=blue,
            urlcolor=blue
            ]{hyperref}

\renewcommand{\IEEEkeywordsname}{Keywords}
\newtheorem{remark}{\bf Remark}

\begin{document}

\title{GenControl:  Generative AI-Driven Autonomous Design of Control Algorithms}

\author{  \vskip 1em Chenggang Cui, \emph{Member, IEEE}, Jiaming Liu, Peifeng Hui, \emph{Student Member, IEEE}, Pengfeng Lin, \emph{Member, IEEE}, Chuanlin Zhang, \emph{Senior Member, IEEE}

}
\maketitle

\begin{abstract}
Modern industrial electronic systems, such as power electronic converters and motor drives, play a pivotal role in applications like new power systems. However, their increasing complexity, nonlinearity, and parameter uncertainties pose significant challenges to the design of high-performance controllers. Traditional model-driven methods heavily rely on system models that are difficult to obtain and maintain accurately, leading to long design cycles, high costs, and insufficient robustness. Existing automation methods also face limitations in structural innovation and synergistic structure-parameter optimization. To address these challenges, this paper proposes a novel autonomous control algorithm design framework driven by Large Language Models (LLMs). The framework employs a bi-level optimization strategy: the upper level utilizes the advanced reasoning and generation capabilities of LLMs for intelligent exploration and iterative optimization of control algorithm structures based on performance feedback; the lower level employs the Particle Swarm Optimization (PSO) algorithm for efficient parameter refinement of a given structure. This approach combines the semantic understanding and knowledge fusion capabilities of LLMs with the search efficiency of traditional optimization algorithms, achieving end-to-end autonomous design from structural conception to parameter tuning. Through a case study on a DC-DC Boost converter (validated by both simulation and hardware experiments), it is demonstrated that the framework can autonomously evolve from a basic template (e.g., standard SMO) to generate a high-performance controller (e.g., Adaptive SMO). This controller effectively meets stringent design specifications (such as fast response, low steady-state error, and strong robustness), significantly enhancing the automation level and efficiency of control design. This work presents a promising new paradigm for the control design of complex industrial electronic systems.
\end{abstract}


\section{Introduction}
In the era of the
new power system, power electronic devices usually serve as interfaces between renewable energy resources and loads \cite{Wang2019_HarmonicStab}. Such devices play a pivotal role in regulating power conversion and distribution. Efficient and effective power transmission and distribution largely depends on control strategies. In general, the design of controllers relies on the parameters of power electronics equipment, accurate mathematical model, and load fluctuations, etc. However, privacy preservation policies and equipment aging issues make it particularly difficult to obtain the accurate parameters of power electronics systems. This fact inevitably complicates the conventional controller design process \cite{Buso2022_DigitalControl}. 

Consequently, traditional model-driven design approaches, while foundational, face significant hurdles. They demand precise mathematical models whose parameters can be difficult to ascertain and maintain accurately over the system's lifetime \cite{Buso2022_DigitalControl, Hu2022_ConverterControlOverview}. This often leads to iterative, time-consuming trial-and-error cycles involving extensive modeling, simulation, tuning, and costly experimental validation \cite{Wu2021_AdvancedControl}. The significant resource demands are exemplified by reports of multi-month commissioning times and daily hardware testing costs potentially reaching tens of thousands of dollars for complex systems like automotive converters \cite{Li2024Controller}. Furthermore, reliance on imperfect models can compromise controller robustness and performance under real-world conditions \cite{Hu2022_ConverterControlOverview}. While model-free control techniques have emerged to mitigate reliance on explicit models, often achieving comparable complexity to PID with superior dynamics \cite{PID_MFC_Complexity_Placeholder}, they frequently still necessitate considerable human expertise for configuration and fine-tuning. This underscores a persistent gap and highlights the critical need for higher levels of automation and autonomy in the controller design workflow to enhance overall efficiency, reduce development costs, and improve resource utilization \cite{MFC_Efficiency_Placeholder1, PID_MFC_Complexity_Placeholder}.

Addressing this need, artificial intelligence and machine learning offer powerful tools for enhancing control system adaptability and design efficiency \cite{Your_ML_Control_Survey_Ref}. Particularly, the transformative advancements in LLMs, characterized by their sophisticated natural language understanding, complex reasoning, and code generation capabilities, present compelling opportunities for automating intricate engineering tasks \cite{LLM_Survey_Placeholder}. Current research actively explores harnessing LLMs, with examples including frameworks like “PE-GPT” aiming to assist in power electronics component selection or parameter suggestion \cite{Lin2024PEGPT}, and higher-level concepts like “Meta-Control” leveraging LLMs for automated control policy synthesis across diverse domains \cite{Meta_Control_Concept_Ref}. Furthermore, the development of LLM-based autonomous agents, capable of planning, tool interaction, and decision-making, signifies a major stride towards realizing systems that can emulate and potentially accelerate the human engineering design process \cite{AutoGen_Placeholder, AgentSurvey_Placeholder}. These convergent trends strongly suggest LLMs can form the core intelligence of frameworks aiming for end-to-end autonomous control system synthesis \cite{Lin2024PEGPT}.

Effectively harnessing the power of LLMs for a complex, multi-faceted task like end-to-end controller design necessitates a structured and coordinated approach. While individual LLM agents show promise, tackling the full design cycle benefits from the principles of Multi-Agent Systems \cite{MAS_Definition_Placeholder}. MAS provide a robust paradigm for decomposing complex problems, assigning specialized roles to different agents, facilitating coordinated action through structured communication, and enabling adaptive system architectures \cite{MAS_Architectures_Placeholder}. Recognizing this synergy, this paper proposes a novel framework that integrates LLM intelligence within a multi-agent architecture specifically designed for the autonomous design and optimization of power electronics controllers. By assigning distinct roles to different agents, orchestrated via CDOPs, our framework leverages both the specific strengths of individual agents and the emergent capabilities arising from their structured collaboration, guided by the LLM's overarching reasoning \cite{Meta_Control_Concept_Ref, LLM_MAS_Collaboration_Placeholder}.

Addressing the aforementioned challenges and opportunities, this paper proposes a novel autonomous control algorithm design framework based on Generative Artificial Intelligence/Large Language Models (GenAI/LLM) and a bi-level optimization strategy. This framework aims to overcome the core bottlenecks in control algorithm design for complex industrial electronic systems, particularly within new power systems, achieving end-to-end automation from structural conception to parameter optimization. The main contributions of this paper include:
\begin{itemize}
    \item Proposing an (implicit) formal representation method for control laws and an extensible primitive library to lay the foundation for structure generation;
    \item Designing an LLM-driven mechanism for structure generation and \emph{Autonomous} iterative optimization, leveraging the reasoning capabilities of LLMs to guide the design process;
    \item Constructing a bi-level optimization framework that effectively decouples and synergizes upper-level LLM-based structure exploration with lower-level Particle Swarm Optimization (PSO)-based parameter refinement;
    \item Developing and validating an end-to-end \emph{Autonomous} control algorithm design workflow oriented towards industrial applications.
\end{itemize}

\section{Control Algorithm Design Problem Formulation}
\label{sec:problem_formulation_final_v14} 

The autonomous design of control algorithms for industrial electronic systems is fundamentally framed as an optimization task. This section is dedicated to providing a formal problem statement, which is essential for the systematic development of the subsequently proposed LLM-driven design framework. The core elements of this problem, including the definition of a controller, the overall search space, the performance objectives, and the operational constraints that guide the synthesis process, will be rigorously defined.
\subsection{Control Algorithm Design Problem}
The control algorithm design problem, denoted as $\mathcal{D}_{\text{problem}}$, is defined by a tuple integrating the essential components that guide controller synthesis:
\begin{equation}
    \mathcal{D}_{\text{problem}} = \{G_{\text{spec}}, M_{\text{plant}}, \mathbb{S}_{\text{set}}, \mathbb{P}, C_{\text{set}}\}
    \label{eq:design_problem_definition_v14}
\end{equation}
where $G_{\text{spec}}$ represents the predefined performance specifications; $M_{\text{plant}}$ is the plant or system model; $\mathbb{S}_{\text{set}}$ is the set of all possible valid control algorithm structures $\mathcal{S}$; $\mathbb{P}$ is the overall search space of candidate controllers $p = \{\mathcal{S}, \mathbf{\theta}\}$, formed from structures in $\mathbb{S}_{\text{set}}$ and their corresponding parameter vectors $\mathbf{\theta}$; and $C_{\text{set}}$ is the set of operational and physical constraints. For iterative design approaches, central to this work, this problem definition is augmented during the solution process by performance feedback from prior evaluations, $P_{\text{feedback}}$, and guiding instructions or prompt templates, $T_{\text{prompt}}$.


\subsection{Objective}
In an iterative design context, $p_k$ may represent the controller candidate at iteration $k$. The overarching goal is to find an optimal controller, $p^*$, from a vast search space of candidate controllers, $\mathbb{P}$. This optimal controller $p^*$ is intended to minimize a performance index, $J(p)$, evaluated for a closed-loop system composed of $M_{\text{plant}}$ and the controller $p$. The entire optimization process is guided by $G_{\text{spec}}$. The objective is formally stated as:
\begin{equation}
    p^* = \arg \min_{p \in \mathbb{P}} J(p) \quad \text{subject to} \quad p \in \mathbb{V}_{\text{valid}}
    \label{eq:opt_objective_final_v8_aligned}
\end{equation}
where $\mathbb{V}_{\text{valid}} \subseteq \mathbb{P}$ represents the subset of controllers satisfying all constraints in $C_{\text{set}}$. These constraints typically encompass requirements for stability, physical realizability, parameter validity, and adherence to critical performance boundaries derived from $G_{\text{spec}}$.
\subsection{Performance Index}
The performance index $J(p)$ quantifies the overall quality of controller $p$. It is often formulated as a weighted sum of $N_{\text{metrics}}$ individual performance metrics $M_i(p)$:
\begin{equation}
    J(p) = \sum_{i=1}^{N_{\text{metrics}}} w_i M_i(p)
    \label{eq:performance_index_weighted_sum_v8_aligned}
\end{equation}
where $w_i$ are non-negative weights. The metrics $M_i(p)$ assess aspects such as tracking precision, control effort, efficiency, robustness, and output quality, with their selection and weighting guided by $G_{\text{spec}}$. The calculation of $J(p)$ for any candidate $p_k$ involves simulation using $M_{\text{plant}}$, and its value serves as a primary component of the performance feedback $P_{\text{feedback}}$.

\subsection{Search Space}

The search space $\mathbb{P}$ is characterized by its hybrid nature. It comprises all possible valid control algorithm structures $\mathcal{S}$ from the set $\mathbb{S}_{\text{set}}$, each paired with a corresponding vector of continuous, real-valued parameters $\mathbf{\theta}$ from its associated parameter space $\Theta_{\mathcal{S}}$. The dimensionality of this parameter vector for a given structure $\mathcal{S}$ is denoted by $d_{\theta}$, and $\Theta_{\mathcal{S}}$ defines the feasible ranges for these $d_{\theta}$ parameters. Thus, every point in $\mathbb{P}$ represents a unique controller instance $\{\mathcal{S}, \mathbf{\theta}\}$. The identification of a globally optimal controller $p^* = \{\mathcal{S}^*, \mathbf{\theta}^*\}$—which minimizes $J(p)$ while ensuring $p^* \in \mathbb{V}_{\text{valid}}$—constitutes a non-trivial joint optimization problem.

In summary, the control algorithm design problem is formulated as identifying $p=\{\mathcal{S}, \mathbf{\theta}\}$ from $\mathbb{P}$ that minimizes $J(p)$ (Eq.~\eqref{eq:performance_index_weighted_sum_v8_aligned}), while satisfying $C_{\text{set}}$ (thus $p \in \mathbb{V}_{\text{valid}}$). This process inherently involves considering $M_{\text{plant}}$ and $G_{\text{spec}}$. For automated, iterative solutions such as the one proposed in this paper, information from $P_{\text{feedback}}$ and guidance via $T_{\text{prompt}}$ are critically employed to navigate the search space. The subsequent sections introduce an LLM-driven framework specifically architected for this task.

\section{Autonomous Design Framework Architecture}
\label{sec:framework_architecture_v3} 

To address the complex control algorithm design problem formalized in Section~\ref{sec:problem_formulation_final_v8}, an autonomous design framework predicated on Large Language Models (LLMs) and a bi-level optimization strategy is proposed herein. This section is dedicated to outlining the comprehensive architecture of this framework and elucidating the functionalities of its constituent modules, as well as its core operational workflow.

\subsection{Framework Philosophy and Architecture}
\label{subsec:framework_philosophy_arch}

\begin{figure}[thpb]
  \centering
  \includegraphics[width=1.0\linewidth]{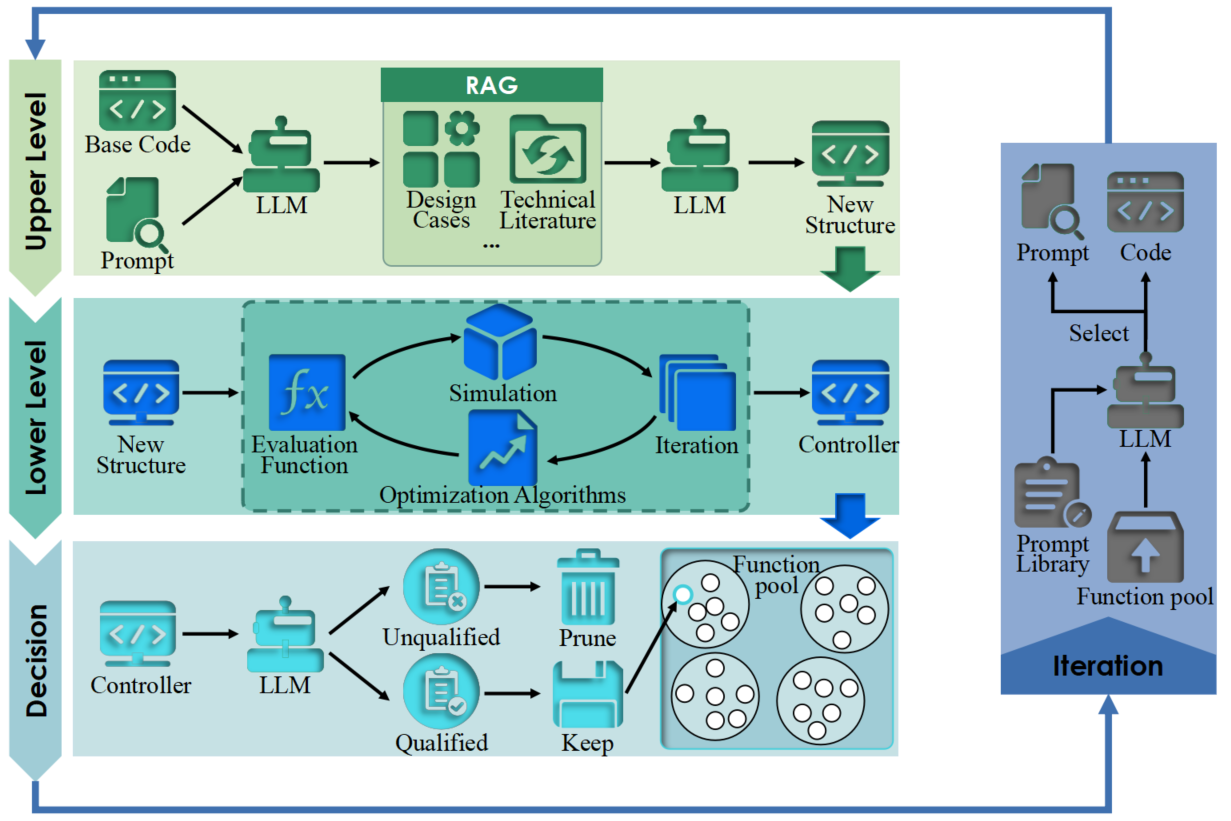} 
  \caption{Overall architecture of the proposed autonomous control algorithm design framework.}
  \label{fig:framework_architecture_main_v3} 
\end{figure}

The proposed framework is engineered to achieve autonomous control design through the synergistic operation of several integral modules. The comprehensive architecture, which underpins the entire autonomous design process, is illustrated in Fig.~\ref{fig:framework_architecture_main_v3}. At its core, the framework leverages the cognitive capabilities of LLMs for high-level structural reasoning and traditional optimization algorithms for efficient parameter tuning, orchestrated within a hierarchical, iterative loop.

\subsection{Module Functionalities}
\label{subsec:module_functionalities}

The autonomous design capability of the framework is realized through the coordinated functionalities of four primary modules:

The \textit{Input Module} serves as the initial interface, responsible for receiving and consolidating all essential data requisite for the design task. This data encompasses a detailed description of the dynamic characteristics of the controlled industrial electronic system, represented as the system model $M_{\text{plant}}$ (which may take the form of mathematical equations or a simulation model). Additionally, the desired closed-loop performance metrics and operational constraints, encapsulated within the performance specifications $G_{\text{spec}}$, are processed by this module. Finally, contextual information and specific instructions intended to guide the LLM's generative and optimization processes are supplied as a prompt template, $T_{\text{prompt}}$.

The \textit{Core Optimization Engine} processes the integrated input information and constitutes the heart of the design iteration. It employs a bi-level optimization structure. At the \textit{Upper Level}, the advanced reasoning and generative capabilities of an LLM are utilized to intelligently explore the space of possible control algorithm structures $\mathcal{S}$ and to generate candidate structures. These candidates are subsequently refined iteratively based on performance feedback. Concurrently, the \textit{Lower Level} of this engine is dedicated to parameter refinement. For any given structure $\mathcal{S}_k$ (where $k$ denotes the iteration index) provided by the upper level, efficient optimization algorithms, such as Particle Swarm Optimization (PSO), are employed to search for a high-quality parameter set $\mathbf{\theta}_k^*$ that optimizes the performance of controller $p_k = \{\mathcal{S}_k, \mathbf{\theta}_k^*\}$.

The \textit{Simulation and Evaluation Module} is tasked with the rigorous performance assessment of each candidate controller $p_k$. This module constructs an appropriate closed-loop simulation environment based on the system model $M_{\text{plant}}$. Simulations are then executed under predefined test scenarios to evaluate the controller's dynamic and steady-state behavior. Following these simulations, the performance index $J(p_k)$ is calculated. Based on this and other diagnostic observations (e.g., constraint satisfaction as per $C_{\text{set}}$, comprehensive performance feedback, $P_{\text{feedback}}$, is generated.

Finally, the \textit{Output Module} is responsible for delivering the consolidated results upon the termination of the optimization process. The primary output is the best-performing optimal controller identified, $p^* = \{\mathcal{S}^*, \mathbf{\theta}^*\}$, representing the culmination of the structural and parametric optimization. In addition, a comprehensive Performance Report is typically generated, which includes detailed evaluation metrics, simulation results, and often a comparative analysis against design specifications or baseline controllers.

\subsection{Iterative Design Workflow}
\label{subsec:iterative_workflow}

The core modules of the framework operate in a tightly coupled, iterative workflow to achieve autonomous design. The process commences with the Input Module gathering all necessary design information ($M_{\text{plant}}$, $G_{\text{spec}}$, $T_{\text{prompt}}$). This information is then fed to the Core Optimization Engine. The upper-level LLM proposes or modifies a control structure $\mathcal{S}_k$, which is passed to the lower level for parameter optimization, yielding $\mathbf{\theta}_k^*$. The resulting controller candidate $p_k = \{\mathcal{S}_k, \mathbf{\theta}_k^*\}$ is then evaluated by the Simulation and Evaluation Module, which computes $J(p_k)$ and generates $P_{\text{feedback}}$.

This crucial feedback $P_{\text{feedback}}$ is then relayed back to the upper-level LLM within the Core Optimization Engine. The LLM analyzes this feedback, informed by $G_{\text{spec}}$ and $T_{\text{prompt}}$, to make intelligent decisions regarding the next iteration. This may involve further refinement of the current structure $\mathcal{S}_k$, generation of a completely new structure, or selection of a different exploratory path. This sequence—structure proposal/modification, parameter optimization, performance evaluation, and feedback-driven LLM decision—forms a closed optimization loop. This iterative process of continuous adjustment and refinement of the controller structure and parameters persists until predefined convergence criteria are satisfied. Such criteria may include the achievement of performance targets outlined in $G_{\text{spec}}$, the exhaustion of a specified design budget (e.g., maximum number of iterations or computational time), or a lack of significant improvement in $J(p)$ over successive iterations. Upon termination, the Output Module presents the final optimized controller $p^*$.




$\Theta_{\mathcal{C}_k}$ to find the parameter combination $\boldsymbol{\theta}_k^*$ that optimizes the performance index $J$. This ensures that the potential of each explored structure is evaluated under its best parameter configuration, answering the question, ``What should the gains/time constants of each component be?''

\section{LLM-Guided Bi-Level Optimization Methodology}
\label{sec:llm_bilevel_opt_methodology}

To effectively address the complex control algorithm design optimization problem, as defined in Section~\ref{sec:problem_formulation_final_v4}, which involves a mixed discrete structure space and continuous parameter space, a bi-level optimization strategy is employed within this framework. This strategy is designed to decouple the exploratory challenge of structural discovery from the precise task of parameter optimization, thereby enabling efficient global search through synergistic interaction between the two hierarchical levels. A notable feature is that the upper-level optimization is driven by a Large Language Model (LLM), which fully leverages its advanced cognitive capabilities to guide the design at the structural level. Concurrently, the lower level utilizes established optimization algorithms, such as Particle Swarm Optimization (PSO), for the refinement of parameters corresponding to a given structure.

\subsection{Rationale for Bi-Level Optimization}
\label{subsec:bilevel_rationale}

The simultaneous optimization of both the control structure $\mathcal{S}$ and its parameters $\mathbf{\theta}$ within the entire mixed-search space $\mathbb{P}$ presents considerable difficulty. The proposed bi-level optimization strategy alleviates this complexity by decomposing the original problem into two interconnected yet distinct subproblems. The \textit{Upper-Level Problem} is concerned with the exploration and selection of the most promising control algorithm structures $\mathcal{S}$ from the set $\mathbb{S}_{\text{set}}$. This constitutes an exploratory task, typically involving discrete or combinatorial optimization. Conversely, the \textit{Lower-Level Problem}, for a specific structure $\mathcal{S}_k$ (where $k$ may denote an iteration index or a candidate identifier) provided by the upper level, is responsible for determining the optimal parameter set $\mathbf{\theta}_k^*$ within its corresponding parameter space $\Theta_{\mathcal{S}_k}$. This latter task represents a more conventional continuous parameter optimization problem. Effective coordination between these levels, facilitated by information transfer and feedback mechanisms, enables the co-evolution of structural exploration and parameter optimization. The outcomes of the lower-level optimization, specifically the optimal performance achievable with a given structure, furnish crucial guidance for the upper level's decisions regarding structural selection or modification. In turn, the upper level directs the lower level on which structures to prioritize for optimization efforts, thereby steering the overall design process.

\subsection{Upper-Level Optimization: LLM-Driven Structural Exploration and Iteration}
\label{subsec:upper_level_llm_driven}

The core innovation of the bi-level optimization strategy resides in the upper-level optimization being driven by an LLM. The role of the LLM can be abstracted as a policy function or mapping, $\pi_{\text{LLM}}$, which guides the search within the structural space $\mathbb{S}_{\text{set}}$. In contrast to traditional structure generation methods that rely on random operators or fixed heuristic rules (e.g., crossover and mutation in Genetic Programming), this framework capitalizes on the advanced cognitive capabilities of LLMs to direct the design and iterative refinement of control structures $\mathcal{S}$. This policy function $\pi_{\text{LLM}}$ accepts current state information, denoted as $\mathcal{X}_k$ (encompassing performance specifications $G_{\text{spec}}$, system model information $M_{\text{plant}}$, historical performance feedback $\{P_{\text{feedback}}^{(i)}\}_{i<k}$, the current controller $p_k = \{\mathcal{S}_k, \mathbf{\theta}_k^*\}$, etc.), and a prompt template $T_{\text{prompt}}$ as input. Its output is the next action, $a_k$, to be taken within the structural space, where $a_k$ could represent the generation of a completely new structure $\mathcal{S}_{k+1}$ or the modification of the current structure $\mathcal{S}_k$ to yield $\mathcal{S}_k'$.
\begin{equation}
    a_k = \pi_{\text{LLM}}(\mathcal{X}_k, T_{\text{prompt}})
    \label{eq:llm_policy_bilevel}
\end{equation}

The LLM's capacity to implement this policy $\pi_{\text{LLM}}$ is predicated on its inherent abilities in contextual understanding and information fusion, knowledge-based reasoning and generation, and goal-oriented decision-making. The LLM can parse and integrate diverse input information, including quantitative performance metrics $J(p_k)$ and qualitative characteristics from $P_{\text{feedback}}^{(k)}$. Guided by $T_{\text{prompt}}$, it can invoke its internal knowledge base concerning control theory and related engineering domains to reason about observed performance deficiencies and propose structural modifications. For instance, a diagnosed steady-state error might lead to an action $a_k = \text{AddIntegralTerm}(\mathcal{S}_k)$. The LLM's decision-making aims to improve $J(p)$ relative to $G_{\text{spec}}$.

The primary function of the upper level is thus the execution of an intelligent search algorithm within $\mathbb{S}_{\text{set}}$, orchestrated by $\pi_{\text{LLM}}$. Based on the performance evaluation $J(p_k)$ obtained from the lower level, the LLM determines the subsequent exploratory action $a_k$. Possible actions include the generation of new structures, typically during initial phases or when structural improvements stagnate; the modification of existing structures, which forms the core of iterative optimization by addressing specific performance issues of $\mathcal{S}_k$; and the selection or pruning of structural branches based on historical performance, to concentrate computational resources. Therefore, the LLM's role significantly transcends that of a mere random generator; it implements an intelligent search strategy that is model-informed (via its internal knowledge) and data-driven (via performance feedback), with the goal of efficiently identifying optimal or near-optimal control algorithm structures. The specific LLM-driven mechanisms, including prompt engineering and the reasoning process, are further elaborated in Section~\ref{sec:llm_mechanisms_v1} (assuming Section V is relabeled).

\subsection{Lower-Level Optimization: Parameter Refinement via PSO}
\label{subsec:lower_level_pso}

For each candidate control algorithm structure $\mathcal{S}_k$ supplied by the upper-level LLM, the objective of the lower-level optimization is to identify the parameter vector $\mathbf{\theta}_k^*$ within its corresponding parameter space $\Theta_{\mathcal{S}_k}$ that optimizes the performance index $J(p)$. This is formulated as a standard parameter optimization subproblem:
\begin{equation}
    \mathbf{\theta}_k^* = \arg \min_{\mathbf{\theta} \in \Theta_{\mathcal{S}_k}} J(\{\mathcal{S}_k, \mathbf{\theta}\})
    \label{eq:lower_level_opt_bilevel}
\end{equation}

Given that the parameter space may exhibit multiple local optima and inter-parameter couplings, this framework predominantly employs the Particle Swarm Optimization (PSO) algorithm as the optimizer for this lower level. PSO is a well-established, population-based global optimization algorithm recognized for its straightforward implementation, relatively rapid convergence, and independence from gradient information of the objective function. These characteristics render it suitable for handling control parameter optimization problems that often present complex nonlinear behaviors. Alternative efficient optimization algorithms, such as Genetic Algorithms (GA) or Differential Evolution (DE), could also be selected depending on specific problem characteristics. The core of this lower-level optimization lies in the design of the fitness function, which is the calculation of the performance index $J(p)$ as detailed in Section~\ref{sec:problem_formulation_final_v4}. This function must accurately and comprehensively reflect practical performance requirements, typically involving a quantitative assessment of multiple aspects and the handling of constraints, often via penalty terms.

\subsection{Inter-Level Coordination Mechanism}
\label{subsec:interlevel_coordination}

The upper and lower optimization levels interact through a well-defined coordination mechanism, establishing an iterative optimization loop. This process begins with \textit{structure transmission}, where the upper-level LLM, guided by its policy $\pi_{\text{LLM}}$, outputs an action $a_k$. This action leads to the generation or selection of one or more candidate structures $\mathcal{S}_k$, which are then transmitted to the lower level. Subsequently, \textit{parameter optimization and evaluation} occur at the lower level. For each received structure $\mathcal{S}_k$, the PSO algorithm (or an alternative optimizer) solves Eq.~\eqref{eq:lower_level_opt_bilevel} to find the optimal parameters $\mathbf{\theta}_k^*$ and calculates the corresponding optimal performance index $J(p_k)$, where $p_k = \{\mathcal{S}_k, \mathbf{\theta}_k^*\}$.

Following evaluation, \textit{performance feedback} is generated. The lower level returns the optimized controller $p_k$, its performance $J(p_k)$, and potentially other diagnostic information (e.g., simulation curve features, constraint violation status), all packaged as $P_{\text{feedback}}^{(k)}$. This feedback is relayed to the upper-level LLM, enabling an update of its state information $\mathcal{X}_{k+1}$. Based on this updated state and the prompt $T_{\text{prompt}}$, the LLM then makes an \textit{LLM decision and initiates the next iteration}. It determines the subsequent action $a_{k+1}$ via its policy $\pi_{\text{LLM}}$. If predefined termination conditions are met (e.g., $J(p_k)$ achieves the targets in $G_{\text{spec}}$, or optimization progress stagnates), $p_k$ is outputted as the final result. Otherwise, action $a_{k+1}$ is executed, leading to a new structure or modification, and the cycle (referred to as the \textit{optimization loop}) repeats from the structure transmission step. This loop continues until stopping criteria, such as a maximum number of iterations $K_{\text{max}}$, exhaustion of computational budget, or convergence of the performance index, are satisfied. Through this bi-level synergistic optimization approach, the framework is designed to effectively navigate the vast search space, balancing the breadth of structural exploration with the depth of parameter optimization, ultimately to achieve the autonomous and efficient design of control algorithms.

\section{LLM-Powered Mechanisms for Iterative Design Optimization}
\label{sec:llm_mechanisms_v1}

The efficacy of the bi-level optimization strategy is heavily contingent upon the intelligence embodied by the upper-level optimizer in navigating the structural search space. A core innovation of this framework is the utilization of a Large Language Model (LLM) as the driving engine for this upper-level optimization. By leveraging its unique cognitive capabilities, the LLM is employed to steer the iterative design process of the control algorithm. This section elaborates on the pivotal roles assumed by the LLM within this optimization loop and delineates its fundamental operational mechanisms.

\subsection{The LLM's Function as an Intelligent Decision Unit in the Optimization Loop}
\label{subsec:llm_role_refined}

Within the proposed autonomous design framework, the LLM transcends the role of a mere structure generator to serve as the \textit{central intelligent decision-making unit} governing the entire iterative optimization loop. Its principal responsibility lies in the processing of performance feedback, $P_{\text{feedback}}$, which is received from the lower-level parameter optimization and subsequent simulation evaluation modules. Based on an analysis of this feedback, and by drawing upon its internal knowledge base, the LLM formulates targeted modification suggestions, denoted as $\Delta p$. These suggestions are instrumental in guiding the controller design process towards the continuous satisfaction of predefined performance specifications, $G_{\text{spec}}$.

The introduction of the LLM marks a fundamental departure from traditional optimization methods that typically rely on predefined rules, stochastic operations, or purely mathematical search techniques. The LLM exhibits capabilities that extend significantly beyond these conventional approaches. Firstly, it is capable of \textit{operating beyond mere syntax}; the LLM not only generates syntactically correct control structures $\mathcal{S}$ but also demonstrates an understanding of the semantics underlying various control strategies. This allows it to effectively link observed performance characteristics to fundamental control principles and the functional roles of different structural components. Secondly, the LLM adeptly \textit{leverages unstructured knowledge}, drawing upon the vast corpus of information assimilated during its training. This includes extensive knowledge of control theory, signal processing, and established engineering practices across diverse application domains, all without necessitating explicit formal encoding of this knowledge. Thirdly, the LLM is proficient in \textit{performing complex reasoning}. It can undertake multi-step inferential processes to analyze the root causes of suboptimal performance and, subsequently, to propose logically sound and coherent improvement schemes, rather than resorting to simplistic local adjustments. This sophisticated capability, founded on deep understanding and nuanced reasoning, empowers the LLM to navigate the complex, joint structure-parameter search space $\mathbb{P}$ with enhanced intelligence and efficiency.

\subsection{An Abstract Model of LLM-Guided Design Mapping}
\label{subsec:llm_mapping_refined}

The guiding role of the LLM within the optimization loop can be abstractly characterized by a mapping function, $\Gamma_{\text{LLM}}$. This function processes current system state information as input and yields a suggested modification action $\Delta p$ as output:
\begin{equation}
    \Delta p = \Gamma_{\text{LLM}}(p_k, G_{\text{spec}}, M_{\text{plant}}, P_{\text{feedback}}, T_{\text{prompt}})
    \label{eq:llm_mapping_refined_eq}
\end{equation}

The inputs to this mapping encompass the current ($k$-th iteration) controller instance $p_k = \{\mathcal{S}_k, \mathbf{\theta}_k^*\}$ (which includes the present structure $\mathcal{S}_k$ and its parameters $\mathbf{\theta}_k^*$ as optimized by the lower level), the predefined performance specifications and constraints $G_{\text{spec}}$, information pertaining to the controlled system model $M_{\text{plant}}$, the performance feedback $P_{\text{feedback}}$ (containing quantitative metrics such as $J(p_k)$ and potentially diagnostic information) obtained from the simulation evaluation module, and the prompt template $T_{\text{prompt}}$, which is designed to steer the LLM towards generating specific types of outputs. The output, $\Delta p$, represents the modification suggestion proposed by the LLM. This suggestion primarily pertains to structural modifications, such as proposing a new or altered structure $\mathcal{S}'$, which subsequently leads to a new controller instance $p' = \{\mathcal{S}', \mathbf{\theta}'\}$ (where $\mathbf{\theta}'$ necessitates re-optimization by the lower level). The precise implementation of this mapping $\Gamma_{\text{LLM}}$ is contingent upon the LLM's internal reasoning processes and is strategically guided by carefully engineered prompts.

\subsection{LLM-Driven Operational Mechanisms}
\label{subsec:llm_key_mechanisms_refined}

The LLM fulfills its guiding role in the optimization loop through the coordinated action of several key operational mechanisms.

A foundational step is \textit{Performance Feedback Interpretation}. The LLM receives the quantitative performance index $J(p_k)$ and potentially richer, more nuanced information embedded within $P_{\text{feedback}}$, such as textual descriptions or data patterns indicating issues like excessive overshoot, persistent steady-state error, or undue sensitivity to specific disturbances. By leveraging its powerful natural language understanding and pattern recognition capabilities, the LLM conducts an in-depth analysis of this feedback. This analysis involves an assessment of the extent to which the current controller $p_k$ meets the performance specifications $G_{\text{spec}}$, a diagnosis of key factors or bottlenecks contributing to suboptimal performance (e.g., inferring insufficient system damping from response curves), and an attempt to localize these diagnosed problems by correlating them with specific parts of the controller structure $\mathcal{S}_k$ or potentially suboptimal parameter values $\mathbf{\theta}_k^*$. This interpretive process establishes the necessary foundation for the subsequent formulation of targeted modification suggestions.

Following interpretation, a critical mechanism is the \textit{Generation of Structural Modifications}, denoted as $\Delta \mathcal{S}$, which provides outer-loop guidance. When performance feedback indicates that the current structure $\mathcal{S}_k$ possesses inherent limitations that cannot be overcome by mere parameter tuning, the LLM initiates this structural modification mechanism. This corresponds to the component of the mapping $\Gamma_{\text{LLM}}$ that yields structural changes, $\Gamma_{\text{LLM},\mathcal{S}} \subseteq \Gamma_{\text{LLM}}$. Based on its interpretation of the performance feedback and by consulting its internal control knowledge base, the LLM generates suggestions for structural modifications $\Delta \mathcal{S}$ or proposes entirely new structures $\mathcal{S}'$. This process can be knowledge-driven; for instance, a diagnosed steady-state error might prompt the LLM to suggest the addition of an integral term at an appropriate point within $\mathcal{S}_k$, or slow response speed might lead to suggestions for introducing derivative or feedforward paths. Furthermore, the LLM may engage in exploratory generation, proposing non-traditional yet potentially promising structural variants based on its exposure to analogous engineering problems, thereby facilitating effective exploration of the structural space $\mathbb{S}_{\text{set}}$. The nature of these generated structures can also be influenced by prompt guidance, whereby $T_{\text{prompt}}$ directs the LLM to produce structures of specific types or those adhering to particular complexity constraints. This mechanism is pivotal for enabling the framework to escape local optima in the structural domain and to explore a broader design landscape, which is essential for the discovery of novel and high-performance control laws.

\begin{remark}[Paradigm Shift and Attendant Advantages]
The LLM-driven iterative optimization mechanism described herein represents a significant \textit{paradigm shift} in the domain of control algorithm design. Its principal advantages are rooted in its capacity for \textit{semantic understanding}; the LLM not only processes symbolic representations but also comprehends the underlying meaning and objectives of the control strategy, thereby rendering the optimization process more intelligent and goal-oriented. Concurrently, it exhibits an ability to \textit{fuse domain knowledge}, naturally leveraging the extensive control theory and engineering practice knowledge embedded within its training data, without the need for complex knowledge representation and explicit encoding processes. Moreover, the generative and reasoning capabilities inherent in the LLM significantly enhance its \textit{exploration ability}. This allows for the exploration of more diverse and even counter-intuitive control structures, thereby increasing the likelihood of discovering breakthrough designs, particularly for complex nonlinear systems that pose challenges for traditional analytical design methods. Ultimately, by emulating the analytical, diagnostic, and decision-making processes characteristic of human experts, the LLM substantially elevates the level of \textit{automation and efficiency} within the control design workflow.
\end{remark}

\subsection{Prompt Design for Control Algorithm Structure Evolution} 
\label{ssubsec:prompt_design_evolution}
The efficacy of LLM-driven structural modification heavily relies on the design of the prompt template ($T_{\text{template}}$). Effective prompts are crucial for translating high-level performance diagnostics into actionable structural changes. Our approach involves a library of prompt templates, each tailored to address specific performance deficiencies or to explore certain structural motifs. For instance, if $P_{\text{feedback}}$ indicates a significant steady-state error, a prompt template might be selected or generated that specifically instructs the LLM: ``Given the current controller structure $\mathcal{C}_k$ and its parameters $\boldsymbol{\theta}_k^*$, which results in a steady-state error of $e_{ss}$, propose a modification to $\mathcal{C}_k$ by incorporating an integral action to eliminate this error. Output the modified structure $\mathcal{C}'$.'' Similarly, prompts can guide the LLM to introduce derivative actions for faster transient response, or to explore non-linear elements if linear structures prove insufficient. The selection or generation of the appropriate $T_{\text{template}}$ itself can be a meta-level decision, potentially also informed by the LLM based on a broader understanding of the design trajectory and historical performance. This structured yet flexible prompting strategy allows the framework to systematically guide the LLM's generative capabilities towards promising regions of the structural search space $\mathfrak{C}$.

\section{Case Study and Results}
\label{sec:simulation_and_results}

To validate the effectiveness and practicality of the proposed LLM-based autonomous control algorithm design framework, a comprehensive simulation case study targeting output voltage regulation in a DC-DC Boost converter was designed and implemented. This section details the simulation model, design objectives, and the results of the autonomous design process.

\subsection{Simulation Setup and Design Objectives}
\label{subsec:simulation_setup}

This case study selects the DC-DC Boost converter as the plant under control, modeled within a simulation environment (e.g., Modelica/Simulink). Its nonlinear characteristics pose certain challenges for controller design. The main nominal parameters of the system model are detailed in Table~\ref{tab:system_parameters}.

To comprehensively evaluate the dynamic performance of the controller, step changes in the load resistance were introduced in the simulation: the load changes from $50\,\Omega$ to $100\,\Omega$ at $t=0.25$\,s, and returns to $50\,\Omega$ at $t=0.5$\,s.

The core performance objectives $G$ for the controller design were set as follows:
\begin{itemize}
    \item Maximum peak overshoot ($M_p$): Required to be less than 5\%.
    \item Steady-state voltage error ($e_{\text{ss}}$): Required to be less than 2\% of the rated output voltage (i.e., less than 2\,V).
\end{itemize}

The autonomous design process was initiated by providing the LLM framework with a structured prompt encapsulating the design task:

\textit{Design a controller for a simulated DC-DC boost converter with the following specifications:}
\begin{itemize}
    \item \textbf{Device Type:} DC-DC Boost Converter
    \item \textbf{Component Parameters:} Load Resistance = $50\,\Omega$, Inductance = 1\,mH, Capacitance = $1100\,\mu$F, Switching frequency = 20\,kHz.
    \item \textbf{Operating Condition:} Input voltage = 50\,V, Target output voltage = 100\,V.
    \item \textbf{Control Objective:} The overshoot must be less than 5\%, and the steady-state error must be less than 2\%.
\end{itemize}

\begin{table}[htbp]
\centering
\caption{Nominal Parameters of the Simulated DC-DC Boost Converter}
\label{tab:system_parameters}
\renewcommand{\arraystretch}{1.5}
\begin{tabular}{|>{\arraybackslash}m{4cm}|>{\centering\arraybackslash}m{1.5cm}|>{\arraybackslash}m{2cm}|}
\hline
\textbf{Parameter} & \textbf{Symbol} & \textbf{Value} \\
\hline
Input Voltage & $V_{\text{in}}$ & 50~V \\
\hline
Output Voltage (Rated) & $V_{\text{out}}$ & 100~V \\
\hline
Inductance & $L$ & 1~mH \\
\hline
Capacitance & $C$ & $1100\,\mu$F \\
\hline
Nominal Load Resistance & $R_{\text{load}}$ & $50\,\Omega$ \\
\hline
Switching Frequency & $f_{\text{sw}}$ & 20~kHz \\
\hline
\end{tabular}
\end{table}

\subsection{Autonomous Design and Optimization Process}
\label{subsec:design_process_and_convergence}

The framework achieved the final high-performance controller not instantaneously, but through multiple rounds of autonomous optimization. This process is characterized by its \textit{iterative optimization}, where the LLM continuously adjusts and optimizes the control strategy based on performance feedback from each simulation run. This feedback-driven structural modification allows the framework to address issues like chattering, steady-state error, or poor dynamic response by making targeted changes to the controller topology.

\begin{figure}[htbp]
    \centering
    \includegraphics[width=0.8\linewidth]{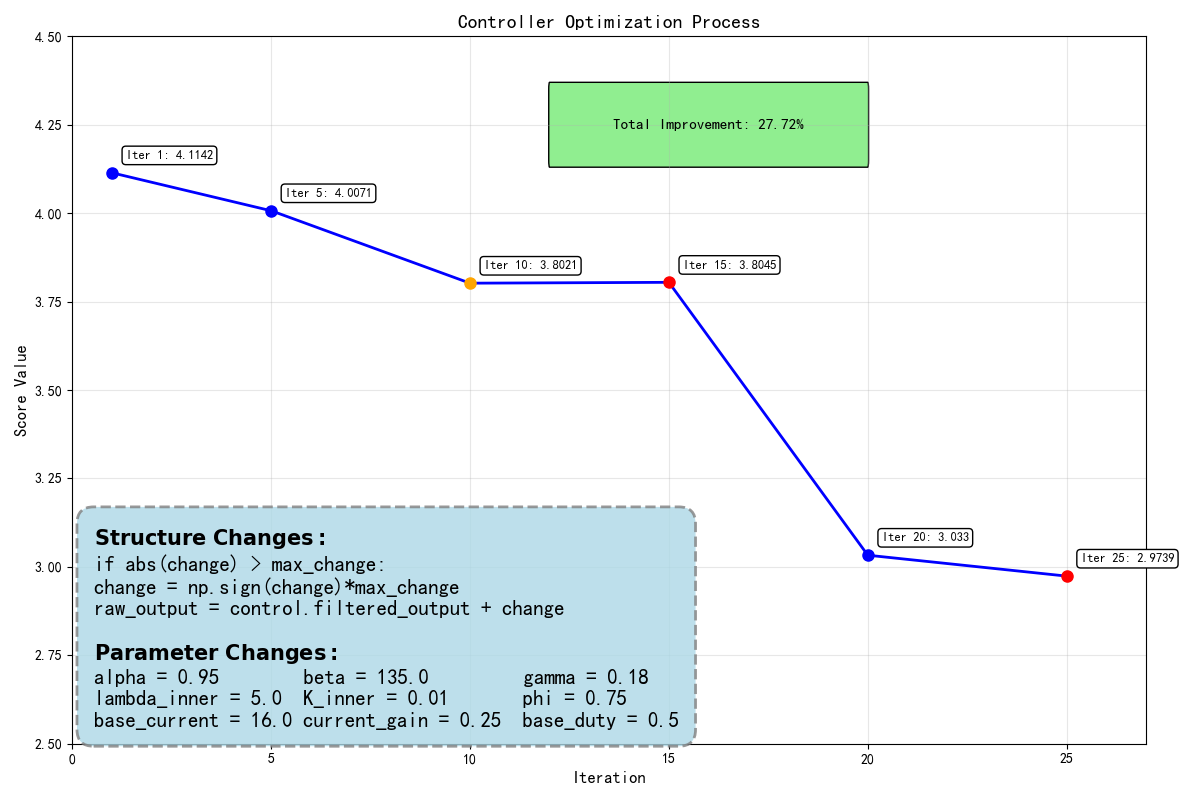}
    \caption{Conceptual illustration of the objective function $J(p)$ convergence during the LLM-driven iterative optimization process.}
    \label{fig:convergence_curve}
\end{figure}

To quantitatively assess this process, we tracked the variation of the objective function $J(p)$ throughout the iterations. As illustrated in Fig.~\ref{fig:convergence_curve}, the objective function value exhibits a significant decreasing trend as the design iterations increase. The process starts with a poorly performing initial controller (high $J(p)$ value) and, through several iterations of structural and parametric refinement guided by the LLM, the performance index rapidly decreases and eventually converges to a low value, signifying that the framework has found a high-performance solution that satisfies the design objectives.

\subsection{Simulation Results and Analysis}
\label{subsec:simulation_results_analysis}

\begin{figure}[htbp]
    \centering
    \includegraphics[width=0.9\linewidth]{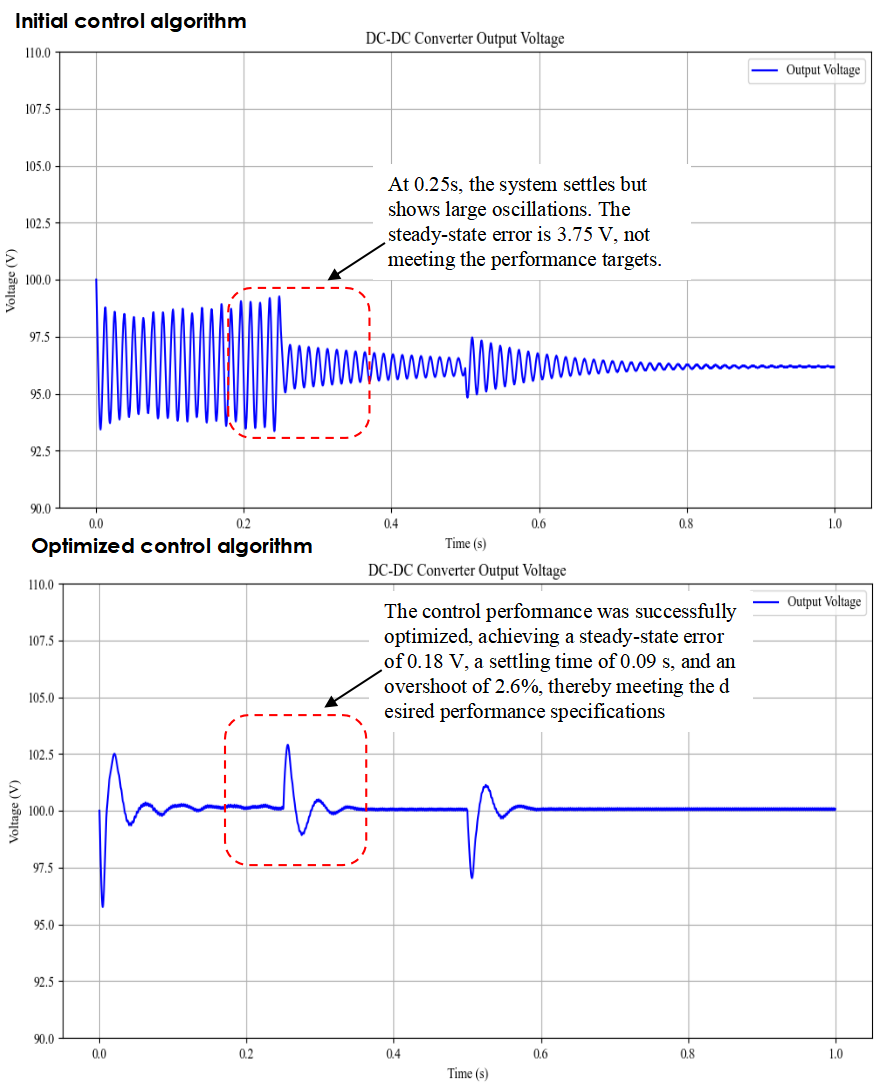}
    \caption{Simulation results of the initial standard SMC controller, showing output voltage response to load changes and the corresponding control signal chattering.}
    \label{fig:smc_results}
\end{figure}

The autonomous design process began with a standard Sliding Mode Control (SMC) structure as the initial baseline controller. Simulation evaluation of this initial controller, as shown in Fig.~\ref{fig:smc_results}, revealed that while the standard SMC has potential, its performance without optimization fell short of the design requirements. Key issues included significant chattering in the control output and a failure to fully meet the specified transient and steady-state performance metrics.

Through the autonomous iterative design process, the framework evolved the initial structure into a significantly improved Adaptive Sliding Mode Controller (Adaptive SMC). The LLM, based on performance feedback, introduced enhancements such as a boundary layer technique to suppress chattering and an adaptive law to improve robustness.

The simulation results of the final optimized Adaptive SMC demonstrate its excellent performance. It achieved minimal steady-state voltage error and exhibited a fast, stable dynamic response with small overshoot under load step disturbances, fully meeting all predefined design objectives. Compared to the initial standard SMC, the final controller's output signal was much smoother, with chattering significantly suppressed. This outcome validates the proposed framework's capability to autonomously evolve a basic controller into an advanced, high-performance solution for a complex control problem entirely within a simulation environment.

\section{Conclusion and Future Work}
\label{sec:conclusion_future}

\subsection{Conclusion}
\label{subsec:conclusion}
This paper addressed the challenges faced in designing control algorithms for industrial electronic systems, particularly within new power systems. We proposed and validated a novel \textit{Autonomous} design framework driven by Large Language Models (LLMs) and employing a bi-level optimization strategy. By adeptly combining the advanced reasoning, knowledge fusion, and generation capabilities of LLMs (for upper-level structure exploration and iterative optimization) with efficient optimization algorithms like Particle Swarm Optimization (PSO) (for lower-level parameter refinement), the framework successfully achieves a high degree of automation in the control algorithm design process.

Through a case study on a DC-DC Boost converter, including both simulation and hardware experimental validation, we demonstrated the effectiveness of the proposed framework. It successfully evolved from a basic controller template (e.g., standard Sliding Mode Control, SMO) through an \textit{Autonomous} iterative optimization process. It identified performance bottlenecks and introduced targeted structural improvements (such as adaptive laws and boundary layers), ultimately designing a high-performance controller (e.g., Adaptive SMO) that met all predefined performance specifications. This not only showcases the framework's capability to solve complex control problems but, more importantly, highlights its significant contribution to enhancing the \textit{Autonomy} of control design, effectively reducing reliance on expert experience and opening possibilities for discovering novel control strategies beyond traditional design paradigms.


\subsection{Future Work}
\label{subsec:future_work}
Although this study has yielded positive results, several directions warrant further exploration in the future, particularly within the industrial power field. Future research could explore more advanced multi-objective optimization techniques (such as Pareto optimization) to simultaneously balance potentially conflicting objectives like performance, efficiency, cost, and robustness, rather than relying primarily on the current weighted-sum approach. Concurrently, hardware constraints and implementation issues, such as computational resource limitations, sensor/actuator precision, and code complexity, should be more explicitly considered during the optimization process to generate controllers that are more readily deployable in practice. Enhancing the interpretability and safety of the generated controller structures and developing formal verification methods are crucial for industrial applications.

Furthermore, applying the framework to a wider variety of more complex industrial electronic systems and validating its performance through extensive physical system verification (including Hardware-in-the-Loop testing and real-world experiments) will be necessary steps. Exploring the extension of the framework towards online adaptation and learning capabilities to cope with unknown environmental changes during operation, as well as integrating data-driven methods to reduce dependence on precise system models and enable learning directly from operational data, are also highly valuable research directions. In-depth investigation into these areas is expected to further enhance the performance, reliability, and applicability of the LLM-driven autonomous control design framework, promoting its widespread adoption in the industrial electronics domain.

\vspace{10pt}
\thispagestyle{plain} 
\end{document}